\documentclass[12pt]{article}

\usepackage{amsmath,amssymb,bm,mathrsfs}
\usepackage{hyperref}
\usepackage{graphicx}

\newcommand{\beq}{\begin{equation}}
\newcommand{\eeq}{\end{equation}}
\newcommand{\beqn}{\begin{eqnarray}}
\newcommand{\eeqn}{\end{eqnarray}}

\newcommand{\beas}{\begin{eqnarray*}}
\newcommand{\eeas}{\end{eqnarray*}}

\newcommand{\bquo}{\begin{quote}}
\newcommand{\enqu}{\end{quote}}

\newcommand{\tN}{\tilde N}

\def\d{\partial}

\def\2{{1\over 2}}
\def\ntwo{${\mathcal N}=2\;$}

\def\ntwot{${\mathcal N}=(2,2)\;$}

\def\ba{\beq\new\begin{array}{c}}
\def\ea{\end{array}\eeq}

\newcommand{\pt}{\partial}

\begin{document}

\begin{flushright}
{FTPI-MINN-16/18, UMN-TH-3528/16 }
\end{flushright}

\vspace{3mm}

\begin{center}
{ \bf \Large Studying Critical  String Emerging from 
\\[2mm]
Non-Abelian Vortex in
 Four Dimensions}

\vspace{6mm}

{\large \bf  P.~Koroteev$^{\,a,b}$,  M.~Shifman$^{\,c}$ and \bf A.~Yung$^{\,\,c,d,e}$}

$^a${\it  Perimeter Institute for Theoretical Physics,
Waterloo, ON N2L2Y5, Canada}\\
$^b${\it  Department of Mathematics, University of California,
Davis, CA 95616}\\
$^c${\it  William I. Fine Theoretical Physics Institute,
University of Minnesota,
Minneapolis, MN 55455, USA}\\
$^{d}${\it National Research Center ``Kurchatov Institute'', 
Petersburg Nuclear Physics Institute, Gatchina, St. Petersburg
188300, Russia}\\
$^{e}${\it  St. Petersburg State University,
 Universitetskaya nab., St. Petersburg 199034, Russia}
\end{center}


\vspace{1mm}
\begin{center}

{\bf \large Abstract}

\end{center}

Recently a special vortex string was found \cite{orixxx} in a class of soliton vortices supported in four-dimensional Yang-Mills theories
that under certain conditions can become infinitely thin and can be interpreted as a critical ten-dimensional string.
The appropriate bulk  Yang-Mills theory
has the $U(2)$ gauge group and the Fayet-Iliopoulos term.
It supports semilocal non-Abelian vortices with the world-sheet theory for orientational and size 
moduli described by the weighted $CP(2,2)$ model. The full target space is $\mathbb{R}^4\times Y_6$
where $Y_6$ is a non-compact Calabi-Yau space.

We study the above vortex string from the standpoint of string theory, focusing on the massless 
states in four dimensions. 
In the generic case all massless modes are non-normalizable, hence,  no massless gravitons 
or vector fields are predicted in the physical spectrum. However, at the selfdual point 
(at strong coupling) weighted $CP(2,2)$ admits deformation of the complex structure,
resulting in a single massless hypermultiplet in the bulk. We interpret it as a composite ``baryon."

\newpage

{\em Introduction.}---Studies of non-Abelian vortex solitons\,\footnote{The non-Abelian
 vortex strings are understood as strings 
carrying non-Abelian moduli on their world sheet, in addition to translational moduli.} \cite{HT1,ABEKY,SYmon,HT2}
supported by some four-dimensional Yang-Mills
theories with ${\mathcal N}=2$ supersymmetry, 
resulted in identification of a bulk theory which under several conditions gives rise
to a critical ten-dimensional string \cite{orixxx}. The appropriate four-dimensional  Yang-Mills theory
has the $U(2)$ gauge group, the Fayet-Iliopoulos term $\xi$ and four matter hypermultiplets.
Its non-Abelian sector would be conformal if it were not for the parameter $\xi$.
It supports semilocal non-Abelian vortices with the world-sheet theory for orientational and size 
moduli
described by the weighted $CP(2,2)$ model.\footnote{A more accurate mathematical term
is the two-dimensional sigma model with the target space $\mathcal{O}(-1)^{\oplus(2)}_{\mathbb{CP}^1}$. 
For brevity we will use $WCP(2,2)$.} The  target space is a six-dimensional
non-compact Calabi-Yau manifold $Y_6$, namely, the resolved conifold. Including the translational moduli with the
$\mathbb{R}^4$ target space one obtains a {\em bona fide} critical string, a 
seemingly promising discovery of  \cite{orixxx}. 

In this paper we explore the spectrum of the massless modes of the above critical closed string
theory. In \cite{orixxx} a hypothesis was formulated regarding parameters of the bulk theory and
the corresponding world sheet model necessary to make the thickness of the
vortex string vanish at strong coupling, which is required in order to neglect higher 
derivative corrections in the world sheet theory on the vortex.
Here, using duality we derive an exact formula for the relation between  the two-dimensional (2D)
coupling  $\beta$  and four-dimensional (4D) gauge coupling $g^2$ of \ntwo SQCD.
Moreover, we identify  the critical point $\beta_*=0$ at which the vortex string 
can become infinitely thin.
At this point the resolved conifold mentioned above 
becomes
a singular conifold. As the only 4D massless mode of the string which emerges at $\beta_*=0$
we identify a single matter hypermultiplet associated with the deformation of the complex structure 
of the conifold. 
Other states arising from the ten-dimensional graviton are not dynamical
in four dimensions. In particular 4D graviton and unwanted vector multiplets are absent.
This is due to non-compactness of the  Calabi-Yau manifold we deal with and 
non-normalizability of the corresponding modes.

Then we discuss the question of how the states seen in the bulk theory at weak coupling are related to
what we obtain from the string theory at strong coupling. In particular we interpret the
 hypermultiplet associated with the deformation of the complex structure  of the conifold
as a monopole-monopole baryon.

\vspace{5mm}

{\em World sheet model.}---The basic bulk theory which supports the string under investigation
is described in detail in \cite{1}. Let us briefly review the model emerging on its world sheet.

The translational moduli fields (they decouple from other moduli)
 in the Polyakov formulation \cite{P81} are given by the action
\beq
S_{\rm 0} = \frac{T}{2}\,\int d^2 \sigma \sqrt{h}\, 
h^{\alpha\beta}\d_{\alpha}x^{\mu}\,\d_{\beta}x_{\mu}
+\mbox{fermions}\,,
\label{s0}
\eeq
where $\sigma^{\alpha}$ ($\alpha=1,2$) are the world-sheet coordinates, $x^{\mu}$ ($\mu=1,...,4$) 
describe the $\mathbb{R}^4$ part  of the string
world sheet and $h={\rm det}\,(h_{\alpha\beta})$, where $h_{\alpha\beta}$ is the world-sheet metric 
which is understood as an
independent variable. The parameter $T$ stands for the tension which will be discussed below.

In the bulk theory under consideration $N_f=2N=4$, implying that in 
addition to  orientational zero modes of the vortex
string $n^P$ ($P=1,2$), there are   size moduli   
$\rho^K$ ($K=1,2$) \cite{AchVas,HT1,HT2,SYsem,Jsem,SVY}.  

The  gauged formulation of the non-Abelian part is as follows \cite{W93}. One introduces
 the $U(1)$ charges $\pm 1$, namely $+1$ for $n$'s and $-1$ for $\rho$'s, 
\beqn
S_{\rm 1} &=& \int d^2 \sigma \sqrt{h} \left\{ h^{\alpha\beta}\left(
 \tilde{\nabla}_{\alpha}\bar{n}_P\,\nabla_{\beta} \,n^{P} 
 +\nabla_{\alpha}\bar{\rho}_K\,\tilde{\nabla}_{\beta} \,\rho^K\right)
 \right.
\nonumber\\[3mm]
&+&\left.
 \frac{e^2}{2} \left(|n^{P}|^2-|\rho^K|^2 -\beta\right)^2
\right\}+\mbox{fermions}\,,
\label{wcp}
\eeqn
where 
\beq 
\nabla_{\alpha}=\d_{\alpha}-iA_{\alpha}\,, \qquad \tilde{\nabla}_{\alpha}=\d_{\alpha}+iA_{\alpha}
\eeq
 and $A_\alpha$ is an auxiliary gauge field.
 The limit $e^2\to\infty$ is implied. Equation (\ref{wcp}) represents the   $WCP(2,2) $ model.\footnote{Both the orientational and the size moduli
have logarithmically divergent norms, see e.g.  \cite{SYsem}. After an appropriate infrared 
regularization, logarithmically divergent norms  can be absorbed into the definition of 
relevant two-dimensional fields  \cite{SYsem}.
In fact, the world-sheet theory on the semilocal non-Abelian string is 
not exactly the $WCP(N,\tN)$ model \cite{SVY}, there are minor differences 
unimportant for our purposes. The actual theory is called the $zn$ model. We can ignore 
the above differences.} 

The total number of real bosonic degrees of freedom in (\ref{wcp}) is six, 
where we take into account constraint
imposed by $D$-term. Moreover, one U(1) phase is gauged away. These six internal degrees of freedom 
are combined with four translational moduli from (\ref{s0}) to form a ten dimensional space needed 
for a superstring to be critical.

In the semiclassical approximation the coupling constant $\beta$ in (\ref{wcp}) is related to 
the bulk $SU(2)$ gauge coupling 
$g^2$
via 
\beq
\beta= \frac{4\pi}{g^2}\,.
\label{betag}
\eeq
Note that the first  (and the only) coefficient of the beta 
functions  is the same for the bulk and 
world-sheet theories  and equals to zero. This ensures that our world sheet theory is conformal
invariant.

The total world-sheet action is
\beq
S=S_0+S_1\,.
\label{5}
\eeq

\vspace{5mm}

{\em Bulk duality vs world sheet duality.}---Since our vortex string is BPS saturated, 
the tension $T$ in Eq. (\ref{s0})
is given by the exact expression
\beq
T=2\pi\xi
\eeq
where $\xi$ is the Fayet-Iliopoulos parameter of the bulk theory. 

As we know \cite{ArgPlessShapiro,APS}  the bulk theory at hand possesses a strong-weak coupling 
duality\,\footnote{Argyres et al. proved this duality for $\xi =0$. It should allow one to 
study the bulk theory at strong coupling in terms of weakly coupled dual theory  
 at $\xi\neq 0$ too.}
\beq
\tau\to \tau_{D} = -\frac{1}{\tau}\,,\qquad \tau = i\frac{4\pi}{g^2} +\frac{\theta_{4D}}{2\pi}\,.
\label{argy}
\eeq
The bulk duality implies a similar 2D duality which
 manifests itself in the world sheet theory as the interchange of the roles of 
the orientational and size moduli,
\beq
n^P \leftrightarrow \rho^K\,,\quad \mbox{or, equivalently, }\beta \to \beta_D=-\beta\,,
\label{swcd}
\eeq
see Eq. (\ref{wcp}). Equation 
 (\ref{betag}) is valid only semiclassically and shows no sign of the
strong-weak coupling duality (\ref{swcd}). An obvious generalization of (\ref{betag}) which 
possess duality (\ref{swcd}) under (\ref{argy})
is 
\beq
\beta = \frac{4\pi}{g^2} -\frac{g^2}{4\pi}\,.
\label{betagexact}
\eeq
If $g^2 \to 16\pi^2/g^2$ then, obviously, $\beta\to -\beta$ as required by (\ref{swcd}). 
The 4D selfdual point
$g^2=4\pi$ is mapped onto $\beta_* = 0$. 
The selfdual point $\beta = 0$ is a critical point
at which the target space  $WCP(2,2)$, which is the resolved conifold, becomes a singular conifold. 

It was conjectured in  \cite{orixxx} that the non-Abelian vortex string become infinitely thin at 
strong coupling  and can be described by the string action (\ref{5}).
The  condition necessary for the vortex string in the bulk 
theory at hand to become
infinitely thin is that $m^2$, the square of the mass of the bulk Higgsed gauge bosons, 
is much larger than the string tension. At weak coupling
$m^2\sim \xi \,g^2$.
It is natural to assume that  mass $m$ goes to infinity at the selfdual point 
$\beta=0$. 
This remains a hypothesis. 
An example of this behavior is 
\beq
m^2 = \frac{4\pi}{|\beta|}\,\xi\,.
\label{11}
\eeq
 
The expansion in derivatives  of the action on the string world sheet runs in 
powers of $\xi/m^2$, implying that
higher derivatives are irrelevant if $m^2\to \infty$ \cite{orixxx}. Note that since the bulk theory has 
a vacuum manifold, there are massless 
states in the bulk. Most of them are not localized on the string, and therefore are 
irrelevant for the string study. 
The only localized zero modes are translational, orientational and size 
moduli\,\footnote{More exactly they have logarithmically divergent norms, a marginal case.} 
\cite{shy}. Other excitations of the string have excitation energies $\sim m^2$.

%
%

One can complexify the constant $\beta$ in a standard way, by adding the 
topological term in the action (\ref{5}) $
\beta_{\rm compl} =   \beta +i\frac{\theta_{2D}}{2\pi}\,,$
where $\theta_{2D}$ is the two-dimensional theta angle which penetrates from the bulk theory 
\cite{Gorsky:2004ad}.
In \cite{tobep} we will present a complexified version of the relation (\ref{betagexact}).

\vspace{2mm}

{\em Bulk 4D supersymmetry from the critical non-Abelian string.}--- The critical string 
discovered in \cite{orixxx} must lead to
${\mathcal N}=2$ supersymmetric spectrum in our bulk four-dimensional QCD. 
This is {\em a priori} clear
because  our starting basic theory is ${\mathcal N}=2$. The question  is how this symmetry emerges 
from the string world-sheet model. 

Non-Abelian  vortex string  is 1/2 BPS; therefore, out of eight bulk supercharges it preserves four.
These four supercharges form ${\mathcal N} =(2,2)$  on the world sheet which 
is necessary to have \ntwo space-time supersymmetry \cite{Gepner,BDFM}. 
\vspace{2mm}

{\em Type IIA vs IIB.}---Another question to ask is  whether the string theory (\ref{5}) 
belongs to Type IIA or IIB? We started with \ntwo supersymmetric QCD
which is a vector-like theory and preserve 4D parity.  Therefore we expect that 
the closed string spectrum
in this theory should respect 4D parity. 

On the other hand we know that Type IIB string
is a chiral theory and breaks parity while Type IIA string theory 
is left-right symmetric and conserves parity \cite{GSW}. Thus we expect that the string
theory of non-Abelian vortex is of Type IIA. In the subsequent paper \cite{tobep} we confirm this
expectations studying how parity transformation acts on the 2D fermions.

\vspace{2mm}

{\em Spectrum: 4D graviton and vector fields}---As was mentioned above, our target space
is $\mathbb{R}^4\times WCP(2,2)= \mathbb{R}^4\times Y_6$ where $Y_6$ is a non-compact  
Calabi-Yau conifold, which in fact, at $\beta\neq 0$  is a resolved conifold, 
see \cite{NVafa} for a review. 
We will consider the string theory (\ref{5}) at small $\beta$.

Strictly speaking at small $\beta$ the sigma model quantum corrections blow up. In other words,
 we can say that at small $\beta$ the gravity approximation does not work. However, 
 for the massless states, we can do the computations at large $\beta$ where supergravity approximation is valid and 
 then extrapolate to strong coupling. This is the reason why now we will limit ourselves only to massless states.
The latter in the sigma model language corresponds to chiral primary operators. They are protected by
\ntwot world sheet supersymmetry and their masses are not lifted by quantum corrections. However,
kinetic terms (the K\"ahler potentials) can be corrected. 

We will argue that all massless string modes on the {\em resolved}
 conifold
(\ref{wcp}) have infinite norm and thus the 4D-graviton and vector field states decouple.

The massless 10D boson fields of type IIA string theory are graviton, dilaton and 
antisymmetric tensor $B_{MN}$ in the NS-NS sector.~In the R-R sector type IIA string gives rise to
 one-form and three-form.
(Above $M,N=1,...,10$ are 10D indices). We start from massless 10D graviton and study what states it can
produce in four dimensions. In fact, 4D states coming from other massless 10D fields listed above can be recovered
from \ntwo supersymmetry in the bulk, see for example \cite{Louis}. We will follow the standard 
string theory approach well developed for compact Calabi-Yau spaces. The only novel aspect
of the case at hand is that for each 4D state we have to check whether its wave function is normalizable on 
$Y_6$ keeping in mind that this space is non-compact.

Massless 10D graviton is represented by fluctuations of the metric  $\delta G_{MN} = G_{MN} - G_{MN}^{(0)}$,
where $G_{MN}^{(0)}$ is the metric on $\mathbb{R}^4\times Y_6$ which has a block form since $R^4$ and 
$Y_6$ are factorized. Moreover, 
$\delta G_{MN}$ should satisfy the Lichnerowicz equation
\beq
D_A D^A \delta G_{MN} + 2R_{MANB}\delta G^{AB}=0\,.
\label{10DLich}
\eeq
Here $D^A$ and $R_{MANB}$ are the covariant derivative and the Riemann tensor corresponding to 
  the background
metric $ G_{MN}^{(0)}$, where $D_A\delta G_{N}^A -\frac12 D_N\delta G_{A}^A=0$. Given the block form
of  $ G_{MN}^{(0)}$,  only the six-dimensional part $R_{ijkl}$ of $R_{MANB}$ is nonzero while the
operator $ D_A D^A$ is given by $ D_A D^A= \pt_{\mu}\pt^{\mu} + D_i D^i$, where 
the indices $\mu,\nu =1,...,4$ 
and $i,j=1,...,6$
belong to flat 4D space and $Y_6$, respectively, and we use 4D metric with the 
diagonal entries $(-1,\,1,\,1,\,1)$.

Next, we search  for the solutions of (\ref{10DLich})
subject to  the block form of $\delta G_{MN}$,
\beq
\delta G_{\mu\nu}=\delta g_{\mu\nu}(x)\,\phi_6(y), \quad \delta G_{\mu i}=B_{\mu}(x)\,V_i(y), 
\quad \delta G_{ij}=\phi_4(x)\,\delta g_{ij}(y)\,,
\label{factor}
\eeq
where $x_{\mu}$ and $y_i$ are the coordinates in $R^4$ and $Y_6$, respectively. We see that
$\delta g_{\mu\nu}(x)$, $B_{\mu}(x)$
and $\phi_4(x)$ are the 4D graviton,  vector  and  scalar fields, while $\phi_6(y)$, $V_i(y)$ and 
$\delta g_{ij}(y)$
are the fields on $Y_6$. 

\vspace{2mm}

Tor the fields $\delta g_{\mu\nu}(x)$, $B_{\mu}(x)$
and $\phi_4(x)$ to be dynamical in 4D, the fields $\phi_6(y)$, $V_i(y)$ and $\delta g_{ij}(y)$ 
should have finite norm
 over the six-dimensional internal space $Y_6$. Otherwise 4D fields will appear with infinite
kinetic term coefficients, and, hence, will decouple.

Symbolically the Lichnerowicz equation (\ref{10DLich}) can be written as
\beq
(\pt_{\mu}\pt^{\mu} + \Delta_6)\,g_4(x)g_6(y)=0,
\label{symbol}
\eeq
where $\Delta_6$ is the two-derivative operator from (\ref{10DLich}) reduced to $Y_6$, while
$g_4(x)g_6(y)$ symbolically denotes the factorized form in (\ref{factor}). If we expand
$g_6$ in eighenfunctions 
\beq
-\Delta_6 g_6(y)=\lambda_6 g_6(y)
\label{lambda}
\eeq
the  eighenvalues  $\lambda_6$ will play the role of the mass squared of the 4D states. 
Here  we will be  only interested in the $\lambda_6=0$ eigenfunctions. Solutions of the
equation $\Delta_6 g_6(y)=0$ for the Calabi-Yau manifolds
are given by elements of Dolbeault cohomology group $H^{(p,q)}(Y_6)$, where 
$(p,q)$ denotes the numbers of holomorphic and anti-holomorphic indices of the form. The numbers 
of these forms 
$h^{(p,q)}$ are called the Hodge numbers for a given $Y_6$.  Due to the fact that $h^{(0,0)}=1$, 
we can easily find (the only!) solution for the 4D graviton:  it is a constant on $Y_6$. This 
mode is non-normalizable. Hence,  no 4D graviton
emerges from our string.

This is a good news. We started from \ntwo QCD in 
four dimensions without gravity and, therefore, do not expect 4D graviton to appear as a closed 
string state.\footnote{The alternative option that 
massless 4D spin-2 state  has no interpretation in terms of 4D gravity is ruled 
out by Weinberg-Witten theorem \cite{ww}.}

The infinite norm of the graviton wave function on $Y_6$  rules out  
other  4D states of the \ntwo 
gravitational and tensor multiplets: the vector field, the dilaton,  the antisymmetric tensor and two scalars
coming from the 10D three-form.

Now, let us pass to the components of the 10D graviton $\delta G_{\mu i}$ which give rise 
to a vector field in 4D. The very possibility of having vector fields is due to continuous symmetries
of $Y_6$, namely for our $Y_6$  we expect seven Killing vectors,which  obey the equations
\beq
D_iV^m_j +D_jV^m_i =0, \qquad m=1,...,7\,.
\label{Killingeq}
\eeq
For Calabi-Yau this implies the equation $\Delta_6 g_6(y)=0$, which takes the form $D_jD^j V_i^m=0$ and
 is equivalent to the statement that 
that $V_i$ is a covariantly constant vector, $D_jV_i=0$. This is impossible for compact Calabi-Yau
 manifolds with the SU(3) holonomy \cite{GSW}, but
possible in the non-compact case. However it is easy to see that the $V_i^m$ fields produced by 
rotations of the $y_i$ coordinates 
do not fall-off at large $y_i^2$ (where the metric tends to flat). Thus, they 
are non-normalizable, and the associated  4D vector fields $B_{\mu}(x)$ are absent. 

\vspace{2mm}

{\em Deformations of the conifold metric.}---It might seem that our 4D string does not produce 
massless 4D states at all. 
This is a wrong impression. At the selfdual value of $\beta=0$ it does! 

Consider the last option in (\ref{factor}), scalar 4D fields. 
In this case the appropriate  Lichnerowicz equation on $Y_6$ reduces
to   
\beq
D_k D^k \delta g_{ij} + 2R_{ikjl}\delta g^{kl}=0\,.
\label{6DLich}
\eeq
The target space of the sigma model (\ref{wcp}) is defined by the $D$ term condition
\beq
|n^P|^2-|\rho^K|^2 = \beta.
\label{Fterm}
\eeq 
Also, a U(1) phase can be  gauged away. 
We can construct the U(1) gauge-invariant ``mesonic'' variables
\beq
w^{PK}= n^P \rho^K.
\label{w}
\eeq
In terms of these variables the condition (\ref{Fterm}) can be written as
${\rm det}\, w^{PK} =0$, or
\beq
\sum_{\alpha =1}^{4} w_{\alpha}^2 =0,
\label{coni}
\eeq
where $w^{PK}=\sigma_{\alpha}^{PK}w_{\alpha}$, and $\sigma$-matrices are  $(1,-i\tau^a)$, $a=1,2,3$.
Equation (\ref{coni}) define the conifold -- a cone with the section $S_2\times S_3$.
It has the K\"ahler Ricci-flat metric and represents a non-compact Calabi-Yau manifold \cite{Candel,W93,NVafa}. 

At $\beta =0$ the conifold develops a conical singularity, so both $S_2$ and $S_3$  shrink to zero.
The conifold singularity can be smoothed in two different ways: by a deformation 
of the K\"ahler form or by a deformation of the 
complex structure. The first option is called the resolved conifold and amounts to introducing 
the  non-zero
$\beta$ in (\ref{Fterm}). This resolution preserves 
the K\"ahler structure and Ricci-flatness of the metric. 
If we put $\rho^K=0$ in (\ref{wcp}) we get the $CP(1)$ model with the $S_2$ target space
(with the radius $\sqrt{\beta}$). The explicit metric for the resolved conifold can be found in 
\cite{Candel,Zayas,Klebanov}. The resolved conifold has no normalizable zero modes. In particular, 
we will demonstrate in \cite{tobep} that the 4D scalar $\beta$ associated with 
deformation of the K\"ahler form
is not normalizable.

If $\beta=0$ another option exists, namely a deformation 
of the complex structure \cite{NVafa}. 
It   preserves the
K\"ahler  structure and Ricci-flatness  of the conifold and is 
usually referred to as the deformed conifold. 
It  is defined by deformation of Eq.~(\ref{coni}), namely   
\beq
\sum_{\alpha =1}^{4} w_{\alpha}^2 = b\,,
\label{deformedconi}
\eeq
where $b$ is a complex number.
Now  the $S_3$ can not shrink to zero, its minimal size is 
determined by
$b$. The explicit metric on the deformed conifold is written down in \cite{Candel,Ohta,KlebStrass}.
 The parameter $b$ becomes a 4D complex scalar field.
The effective action for  this field is
\beq
S(b) = T\int d^4x \,h_{b}|\pt_{\mu} b|^2,
\label{Sb}
\eeq
where the metric $h_{b}(b)$ is given by the normalization integral over the conifold $Y_6$,
\beq
h_{b} = \int d^6 y \sqrt{g} g^{li}\left(\frac{\pt}{\pt b} g_{ij}\right)
g^{jk}\left(\frac{\pt}{\pt \bar{b}} g_{kl}\right),
\label{hbgen}
\eeq
and $g_{ij}(b)$ is the deformed conifold metric.

We calculated $h_b$  by two different methods, one of them \cite{GukVafaWitt} was 
kindly pointed out to us by Cumrun Vafa. Details of these calculations will be presented 
in \cite{tobep}. Here we only state that the norm of
the $b$ modulus is marginal (i.e. logarithmically divergent), in full accord with the 
analysis of \cite{GukVafaWitt}.
Thus, this scalar mode is localized on the string in the same sense as the orientational 
and size zero modes 
are localized on the vortex-string  solution.
   
   In type IIA superstring the complex scalar associated with deformations of the complex structure of 
   the Calabi-Yau
space enters
as a 4D hypermultiplet. Thus our 4D scalar $b$ is a part of a hypermultiplet. Another complex scalar
$\tilde{b}$ comes from 10D three-form, see \cite{Louis} for a review. Together they form the bosonic
content
 of a 4D \ntwo hypermultiplet. The fields $b$ and $\tilde{b}$ being massless can develop VEVs. Thus, 
we have a new Higgs branch in the bulk which is developed only at the self-dual value of 
coupling constant $g^2=4\pi$. 
The bosonic part of the full effective action for the
 $b$ hypermultiplet to be presented  in \cite{tobep} takes the  form
\beq
S(b) = T\int d^4x \left\{|\pt_{\mu} b|^2 +|\pt_{\mu} \widetilde{b}|^2 \right\}\,
\log{\frac{T^4 L^8}{|b|^2+| \widetilde{b}|^2}}\,,
\label{Sbtb}
\eeq
where $L$ is the  size of $R^4$ introduced as an infrared regularization of logarithmically divergent
norm of $b$-field.

The logarithmic metric in (\ref{Sbtb}) in principle can receive both perturbative and 
non-perturbative quantum corrections. However, for \ntwo  theory the non-renormalization
theorem of \cite{APS} forbids the  dependence of the Higgs branch metric  on the 4D coupling 
constant $g^2$.
Since the 2D coupling $\beta$ is related to $g^2$ we expect that the logarithmic metric in (\ref{Sbtb})
will stay intact.

The presence of the 
``non-perturbatively emergent'' Higgs branch
 at the self-dual point  $g^2=4\pi$ at strong coupling is a  successful test of our picture. 
A hypermultiplet is a BPS state. If it were present in a continuous region of $\tau$ 
at strong coupling it could be continued all the way to the weak coupling domain where its presence would 
contradict the  quasiclassical analysis of bulk QCD \cite{1,tobep}.

\vspace{2mm}

{\em String states interpretation in the bulk.}---To find the place for the massless scalar hypermultiplet
we obtained as the only massless string excitation at the critical point $\beta=0$ let us first 
examine weak coupling
domain $g^2\ll 1$.

As well-known from the previous studies of the vortex-strings at 
weak coupling, non-Abelian vortices confine 
monopoles. The elementary 
monopoles
are junctions of two distinct elementary  non-Abelian strings \cite{T,SYmon,HT2}. As a result
in the bulk theory  we have 
monopole-antimonopole mesons in which monopole and antimonopole are connected by two confining strings.
 For the U(2) gauge group we have also  ``baryons" consisting
of two monopoles, rather than of monopole-antimonope pair. 
The monopoles acquire quantum numbers with respect to the global group $SU(2) \times SU(2)\times U(1)$
of the bulk theory. Indeed,
  in the world sheet
model on the vortex-string confined monopole are seen as  kinks interpolating between two different vacua \cite{T,SYmon,HT2}. These  kinks  are described at strong coupling by the $n^P$ and $\rho^K$ fields \cite{W79,HoVa}
(for $WCP(N,\tN)$ models see \cite{SYtorkink}) and therefore transform in 
the fundamental representations of $SU(2) \times SU(2)$ for $WCP(2,2)$ \footnote{Here we note that global
group $SU(2) \times SU(2)\times U(1)$ is the same for both bulk and world sheet theories \cite{1}}.

As a result, the monopole-antimonopole mesons and baryons can be  either singlets or triplets
of both $SU(2)$ global groups, as well as in the bi-fundamental representations. With respect
to baryonic $U(1)_B$ symmetry which we define as a U(1) factor in the 
global $SU(2) \times SU(2)\times U(1)$, the
 mesons have charges $Q_{B}({\rm meson})=0,1$ while the
baryons can have charges
\beq
Q_{B}({\rm baryon})=0,1,2\,.
\label{Bbaryons}
\eeq
 All the above non-perturbative stringy states are heavy, with mass of the
order of $\sqrt{\xi}$, and can decay into screened quarks, which are lighter, and, eventually, into
massless bi-fundamental screened quarks.

Now we return from weak to strong coupling and go to the  self-dual point $\beta=0$. 
We claim that at this point a new `exotic' Higgs branch opens up which 
is parameterized by the massless hypermultiplet -- the $b$ state,  associated with the deformation of the 
complex structure of the conifold. It can be interpreted as a baryon
constructed of two monopoles.  To this end note that the complex 
parameter $b$ (promoted to a 4D scalar field) is singlet with respect to two $SU(2)$ factors
of the global world-sheet group while its baryonic charge is $Q_{B}=2$ \cite{tobep}.

Since the world sheet and the bulk global groups
are identical we can conclude that our massless $b$ hypermultiplet is a monopole-monopole baryon.

Being massless it is marginally stable at $\beta=0$ and can decay into pair of massless bi-fundamental
quarks in the singlet channel with the same baryon charge $Q_{B}=2$. The 
$b$ hypermultiplet does not exist at non-zero ${\beta}$.

\vspace{2mm}

 {\em Acknowledgments.}---We are very grateful to Igor Klebanov and Cumrum Vafa for very useful correspondence and insights, and 
to Nathan Berkovits, Alexander Gorsky, David Gross, Zohar Komargodski, Andrei Mikhailov and 
Shimon Yankielowicz for 
helpful comments.

The work of M.S. is supported in part by DOE grant DE-SC0011842. 
The work of A.Y. was  supported by William I. Fine Theoretical Physics Institute,   
University of Minnesota,
by Russian Foundation for Basic Research Grant No. 13-02-00042a and by Russian State Grant for
Scientific Schools RSGSS-657512010.2. The work of A.Y. was supported by the Russian Scientific Foundation 
Grant No. 14-22-00281.  P.K. would like to thank W.~Fine Institute for Theoretical Physics at the University of Minnesota for kind hospitality during his visit, where part of his work was done. The research of P.K. was supported in part by the Perimeter Institute for Theoretical Physics. Research at Perimeter Institute is supported by the Government of Canada through Industry Canada and by the Province of Ontario through the Ministry of Economic Development and Innovation.


\begin{thebibliography}{99}


\bibitem{HT1}
A.~Hanany and D.~Tong,
JHEP {\bf 0307}, 037 (2003).

\bibitem{ABEKY}
R.~Auzzi, S.~Bolognesi, J.~Evslin, K.~Konishi and A.~Yung,
Nucl.\ Phys.\ B {\bf 673}, 187 (2003).

\bibitem{SYmon}
M.~Shifman and A.~Yung,
Phys.\ Rev.\ D {\bf 70}, 045004 (2004).

\bibitem{HT2}
A. Hanany and D. Tong,
JHEP {\bf 0404}, 066 (2004).

\bibitem{orixxx} 
  M.~Shifman and A.~Yung,
 {\em Critical String from Non-Abelian Vortex in Four Dimensions,}
  Phys.\ Lett.\ B {\bf 750}, 416 (2015)
  [arXiv:1502.00683 [hep-th]].

\bibitem{1}
For a review and extensive references see e.g. M. Shifman and A. Yung, 
{\sl Supersymmetric Solitons}, Cambridge University Press, 2009.

\bibitem{P81}
A.~Polyakov,
Phys.\ Lett. \ {\bf B103}, 207 (1981).

\bibitem{AchVas}
For a review see e.g. A.~Achucarro and T.~Vachaspati,
  Phys.\ Rept.\  {\bf 327}, 347 (2000).

\bibitem{SYsem}
 M.~Shifman and A.~Yung,
  Phys.\ Rev.\  D {\bf 73}, 125012 (2006).
  
\bibitem{Jsem}
M.~Eto, J.~Evslin, K.~Konishi, G.~Marmorini, et al.,
  Phys.\ Rev.\  D {\bf 76}, 105002 (2007).
  
\bibitem{SVY}
 M.~Shifman, W.~Vinci and A.~Yung,
 ``Effective World-Sheet Theory for Non-Abelian Semilocal Strings in N = 2 Supersymmetric QCD,''
  Phys.\ Rev.\ D {\bf 83}, 125017 (2011);
  [arXiv:1104.2077 [hep-th]].
  P.~Koroteev, M.~Shifman, W.~Vinci and A.~Yung,
  ``Quantum Dynamics of Low-Energy Theory on Semilocal Non-Abelian Strings,''
  Phys.\ Rev.\ D {\bf 84}, 065018 (2011)
  [arXiv:1107.3779 [hep-th]].
  
\bibitem{W93}
E.~Witten,
  ``Phases of N = 2 theories in two dimensions,''
  Nucl.\ Phys.\ B {\bf 403}, 159 (1993).
  [hep-th/9301042].
  
\bibitem{ArgPlessShapiro}
P.~Argyres, M.~R.~Plesser and  A.~Shapere,
{\em The Coulomb Phase of N=2 Supersymmetric QCD}
Phys.\ Rev. \ Lett. {\bf 75}, 1699 (1995).
[hep-th/9505100].

 \bibitem{APS}
P.~Argyres, M.~Plesser and N.~Seiberg,
{\em The Moduli Space of ${\mathcal N}=2$  SUSY QCD and Duality in
${\mathcal N}=1$  SUSY QCD,}
Nucl. Phys. {\bf B471}, 159  (1996).

\bibitem{Gorsky:2004ad} 
  A.~Gorsky, M.~Shifman and A.~Yung,
{\em Non-Abelian Meissner effect in Yang-Mills theories at weak coupling,}
  Phys.\ Rev.\ D {\bf 71}, 045010 (2005)
  [hep-th/0412082].
  
\bibitem{shy}
M.~Shifman and A.~Yung,
{\em Non-Abelian semilocal strings in} ${\mathcal N}=2$ {\em supersymmetric QCD,}
  Phys.\ Rev.\ D {\bf 73}, 125012 (2006)
  [hep-th/0603134].
    
\bibitem{Gepner}
D.~Gepner,
{\em Space-Time Supersymmetry in Compactified String Theory and Superconformal Models,}
Nucl. \ Phys. \ B {\bf 296}, 757 (1988).

\bibitem{BDFM}
T.~Banks, L.~J.~Dixon, D.~Friedan, and E.~J.~Martinec, 
{\em Phenomenology and Conformal Field Theory Or Can String Theory Predict  the Weak Mixing Angle? }
Nucl. \ Phys. \ B  {\bf 299}, 613 (1988).

\bibitem{GSW}
M.~B.~Green, J.~H.~Schwarz and E.~Witten,
{\sl Superstring Theory}
(Cambridge University Press, 1987).

\bibitem{tobep}
P.~Koroteev, M.~Shifman and A.~Yung
{\em Non-Abelian Vortex in Four Dimensions as a Critical String on a Conifold,}
To be published.

 \bibitem{NVafa}
A.~Neitzke and  C.~Vafa,
{\em Topological strings and their physical applications},
arXiv:hep-th/0410178.

 \bibitem{Louis}
J.~Louis
{\em Generalized Calabi-Yau compactifications with $D$-branes and fluxes,}
Fortschr. Phys. {\bf 53}, 770 (2005).

\bibitem{ww}
  S.~Weinberg and E.~Witten,
  ``Limits on Massless Particles,''
  Phys.\ Lett.\ B {\bf 96}, 59 (1980).
  
\bibitem{Candel}
P.~Candelas and X.~C.~ de la Ossa,
{\em Comments on conifolds,}
Nucl. \ Phys. \ {\bf B342}, 246 (1990).

\bibitem{Zayas}
L.~A.~Pando Zayas and  A.~A.~Tseytlin,
{\em 3-branes on resolved conifold,}
 JHEP \ {\bf 0011},  028 (2000)
[arXiv:hep-th/0010088].

\bibitem{Klebanov}
I.~R.~Klebanov and  A.~Murugan,
{\em Gauge/Gravity Duality and Warped Resolved Conifold,}
JHEP \ {\bf 0703}, 042 (2007)
[arXiv:hep-th/0701064].

\bibitem{GukVafaWitt}
S.~Gukov, C.~Vafa and E.~Witten
{\em CFT's from Calabi-Yau four folds, }
Nucl. \ Phys. \ {\bf B584},  69 (2000)
[arXiv:hep-th/0410178].

\bibitem{Ohta}
K.~Ohta and T.~Yokono,
{\em Deformation of Conifold and Intersecting Branes,}
 JHEP \ {\bf 0002}, 023 (2000)
[arXiv:hep-th/9912266]

\bibitem{KlebStrass}
I.~R.~Klebanov and  M.~J.~Strassler,
{\em Supergravity and a Confining Gauge Theory: Duality Cascades and $chi$SB-Resolution of 
Naked Singularities,}
JHEP\ {\bf 0008}, 052 (2000)
[arXiv:hep-th/0007191].

  \bibitem{T}
D.~Tong,
{\em Monopoles in the Higgs phase,}
Phys.\ Rev.\ D {\bf 69}, 065003 (2004).


\bibitem{W79} 
E.~Witten,
{\em Instantons, the Quark Model, and the 1/N Expansion,}
  Nucl.\ Phys.\ B {\bf 149}, 285 (1979).

 \bibitem{HoVa}
  K.~Hori and C.~Vafa,
{\em Mirror symmetry,}
  arXiv:hep-th/0002222.
  
  \bibitem{SYtorkink}
M.~Shifman and A.~Yung,
{\em Non-Abelian Confinement in ${\mathcal N}=2$  Supersymmetric QCD: Duality and Kinks on
 Confining Strings,}
  Phys.\ Rev.\  D {\bf 81}, 085009 (2010)
  [arXiv:1002.0322 [hep-th]].



\end{thebibliography}
\end{document}